\documentclass[twocolumn,superscriptaddress,aps,showpacs,amsmath,footinbib,bibnotes]{revtex4-1}
\pdfoutput=1
\usepackage{graphicx}
\usepackage{amssymb,amsmath}
\usepackage{textcomp}
\usepackage[bold]{hhtensor}
\usepackage{natbib}
\usepackage{lipsum}
\usepackage{bm}
\usepackage{hyperref}
\usepackage{multirow}
\usepackage{float}
\usepackage{amsmath}
\usepackage{amssymb}
\usepackage{stmaryrd}
\usepackage{mathrsfs}
\usepackage[dvipsnames]{xcolor}
\hypersetup{
	colorlinks,
	linkcolor={red!80!black},
	citecolor={Blue},
	urlcolor={MidnightBlue}
}

\begin{document}

\author{Lingyan He}
\affiliation{John A. Paulson School of Engineering and Applied Sciences, Harvard University, Cambridge, Massachusetts 02138, USA}
\affiliation{Department f Physics, Beijing University of Posts and Telecommunications, Beijing, 100876, China}
\author{Mian Zhang}
\affiliation{John A. Paulson School of Engineering and Applied Sciences, Harvard University, Cambridge, Massachusetts 02138, USA}
\affiliation{HyperLight Corporation, 501 Massachusetts Avenue, Cambridge, MA 02139, USA}
\author{Amirhassan Shams-Ansari}
\affiliation{John A. Paulson School of Engineering and Applied Sciences, Harvard University, Cambridge, Massachusetts 02138, USA}
\author{Rongrong Zhu}
\affiliation{John A. Paulson School of Engineering and Applied Sciences, Harvard University, Cambridge, Massachusetts 02138, USA}
\affiliation{The Electromagnetics Academy at Zhejiang University, College of Information Science and Electronic Engineering, Zhejiang University, Hangzhou 310027, China}
\author{Cheng Wang}
\affiliation{John A. Paulson School of Engineering and Applied Sciences, Harvard University, Cambridge, Massachusetts 02138, USA}
\affiliation{Department of Electronic Engineering, City University of Hong Kong, Kowloon, Hong Kong, China}
\author{Marko Loncar}\email{loncar@seas.harvard.edu}
\affiliation{John A. Paulson School of Engineering and Applied Sciences, Harvard University, Cambridge, Massachusetts 02138, USA}
\date{\today}
\title{Low-loss fiber-to-chip interface for lithium niobate photonic integrated circuits}

\begin{abstract}
Integrated lithium niobate (LN) photonic circuits have recently emerged as a promising candidate for advanced photonic functions such as high-speed modulation, nonlinear frequency conversion and frequency comb generation. For practical applications, optical interfaces that feature low fiber-to-chip coupling losses are essential. So far, the fiber-to-chip loss (commonly $>$ 10 dB) dominates the total insertion losses of typical LN photonic integrated circuits, where on-chip propagation losses can be as low as 0.03 - 0.1 dB/cm. Here we experimentally demonstrate a low-loss mode size converter for coupling between a standard lensed fiber and sub-micrometer LN rib waveguides. The coupler consists of two inverse tapers that convert the small optical mode of a rib waveguide into a symmetric guided mode of a LN nanowire, featuring a larger mode area matched to that of a tapered optical fiber. The measured fiber-to-chip coupling loss is lower than 1.7 dB/facet with high fabrication tolerance and repeatability. Our results open door for practical integrated LN photonic circuits efficiently interfaced with optical fibers.
\end{abstract}
\maketitle

Integrated lithium niobate (LN) photonics has been growing rapidly over the past decade, due to the attractive material properties including a large second-order nonlinear susceptibility, a large piezoelectric response, a wide optical transparency window,  a high refractive index \cite{ferraro2013ferroelectric}, and the recently developed thin-film LN-on-insulator (LNOI) nanofabrication technology \cite{RN72}. A wide range of nanophotonic devices including microring resonators \cite{guarino2007electro,RN66, wolf2017cascaded, RN55,krasnokutska2018ultra}, photonic crystal cavities \cite{witmer2016design,RN33} and microdisk resonators \cite{RN27,wang2015high, wu2018lithium} have been realized and are used to achieve advanced optical functionalities on-chip such as electro-optic (EO) modulation \cite{RN26,RN65,RN50,weigel2018bonded,mercante2018thin,he2018high,escale2018extreme,witmer2018high}, second harmonic generation \cite{RN55,RN25,RN69}, supercontinuum generation \cite{RN74}, electro-optic frequency comb \cite{RN20} and Kerr frequency comb generation \cite{RN71,he2018self}. These on-chip functionalities could have profound impact on integrated optical solutions in optical communications \cite{RN25,RN55,wang2015high}, spectroscopy and sensing \cite{RN56}, as well as microwave photonics \cite{RN18,RN75,RN52}.

However, a major challenge for practical applications of integrated LN photonics is the lack of an efficient interface between the micrometer-scale LN devices and optical fibers. For many applications including electro-optic modulators and frequency comb generation, the system performance is directly associated with the total optical power delivered through the chips. For example, integrated LN modulators developed recently feature large bandwidths, low driving voltages and low on-chip losses, but at the same time suffer from large fiber-to-fiber insertion losses in excess of 10 dB \cite{RN50,RN18}. This is due to a large mismatch between the mode sizes and mode indices of optical fibers and nanophotonic LN waveguides. Such a high insertion loss prevents these high-performance modulators to reach their full potential and find practical applications. Increasing the on-chip optical mode size is therefore important not only to enable LN photonic integrated circuits with low insertion losses, but also to improve the alignment tolerance during packaging and assembly.

Here we demonstrate an on-chip lithium niobate bilayer inversely tapered mode size converter and achieve fiber-to-chip coupling losses lower than 1.7 dB/facet (fiber-to-fiber insertion loss of 3.4 dB) at telecom wavelengths through an optimized nanofabrication process. This is a substantial  improvement over recent fiber-chip interfaces in thin-film LN devices, including grating couplers with insertion losses of $\sim$10 dB or higher \cite{RN31,RN23}, and end-fire couplers on a hybrid silicon nitride/LN platform ($\sim$ 6 dB/facet)\cite{RN25}.

\begin{figure}
	\centering
	\includegraphics[angle=0,width=0.5\textwidth]{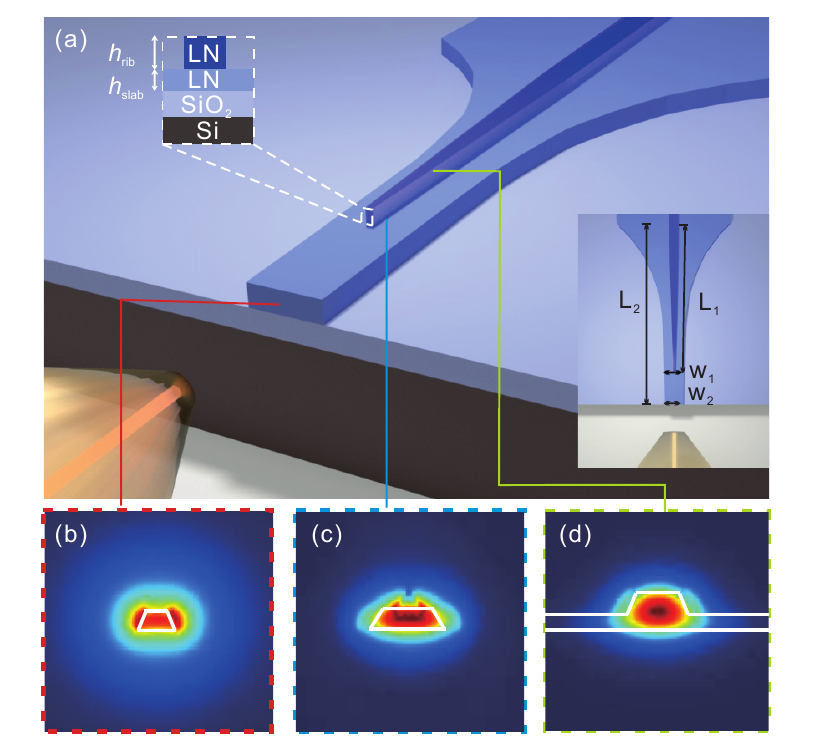}
	
	\caption{\label{fig1}\textbf{Bilayer mode size converter.}\textbf{(a)} Schematic of the bilayer tapered mode size converter. Top-left inset: cross-sectional view of the bilayer inverse taper. Bottom right inset: top-down view of the bilayer inverse taper. $L_1$=120 $\mu$m and $L_2$=450 $\mu$m are the tapering lengths of the top rib waveguide and the bottom slab layer, respectively. $w_1$=1.2 $\mu$m corresponds to the width of the bottom-layer taper in the conversion section and $w_2$=340 nm is the tip width. Note that the widths here are defined as the top widths of the trapezoids. \textbf{(b-d)} Cross-sectional electrical field distribution (Ez) of the TE optical modes at different tapering regions: \textbf{(b)} tip of the bottom slab layer, \textbf{(c)} conversion area where top waveguide ridge just ends, \textbf{(d)} and bilayer tapers \textbf{(d)}. }
\end{figure}

Many integrated LN devices rely on a rib waveguide geometry, with a slab beneath the ridge. The slab geometry is desired to allow the electric field to efficiently propagate through LN with has a relatively large dielectric constant. As a result, conventional single layer inverse taper designs \cite{RN70} are not readily suitable here since tapering the rib portion only will push optical mode to the LN slab resulting in poor coupling efficiencies. Therefore, efficient couplers for rib waveguide structures require a bilayer taper to convert the etched rib waveguide mode to a nearly circularly symmetric output mode. In our approach, this is accomplished by adiabatically tapering the bottom slab layer in addition to the top ridge waveguide. Fig. \ref{fig1}(a) shows the schematic view of our LN mode size converter designed to support a fundamental transverse-electric (TE) optical mode, where in-plane electric fields ($E_z$) could interact through the highest electro-optic/nonlinear-optic tensor component of the x-cut LN thin film we use. The coupling is based on a gradual variation of the waveguide cross section to a certain point such that better size overlap and index match between modes of the input fiber and nanophotonic waveguides are achieved.

\begin{figure}
	\centering
	\includegraphics[angle=0,width=0.5\textwidth]{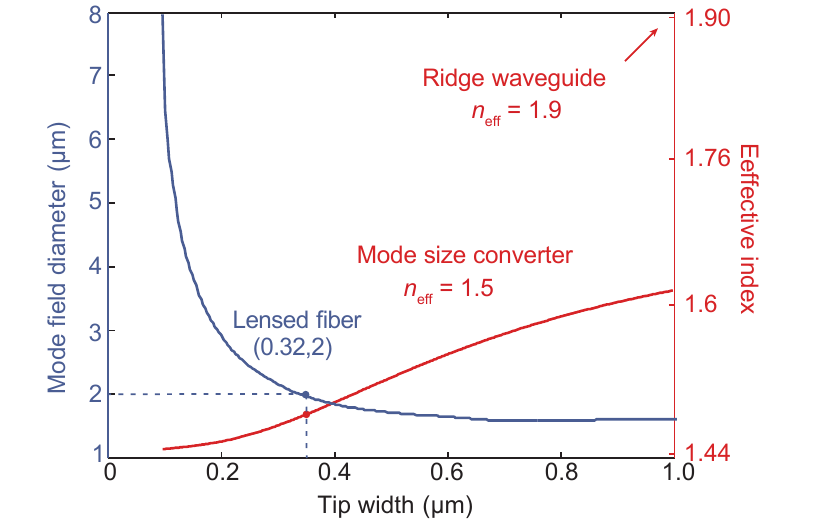}
	
	\caption{\label{fig2}\textbf{Inverse taper mode matching.} Dependence of mode field diameter (blue curve) and effective refractive index (red curve) on the tip widths of the slab layer, respectively. Inset: the optical profile of the ridge waveguide without mode size converter.}
	
\end{figure}

Finite-difference time domain (FDTD) simulations are used to design an adiabatic coupler \cite{RN76} that can match the optical mode of a lensed single mode fiber (SMF) with a mode field diameter (MFD) of $\sim$ 2 $\mu$m in air ($n_\textrm{air} $= 1), to the mode of a LN rib waveguide, with MFD $\sim$ 1.6 $\mu$m and SiO$_2 $cladding ($n_\textrm{LN}=2.2$, $n_\textrm{SiO$_2$}=1.45$). To accomplish this, we utilize gradual tapering of both rib and slab portions of the waveguide following an exponential function. The top (rib) taper evolves from a nominal LN waveguide width of 800 nm down to a 30-nm tip with a height $h_\textrm{rib}$ of 350 nm over a 120-$\mu$m length, which ensures single-mode propagation along the waveguide taper. The bottom (slab) layer taper narrows down laterally from 6 $\mu$m to a tip of 340 nm wide with a slab height $h_\textrm{slab}$ of 250 nm to match the optical mode of fiber. The optical mode profiles of the fundamental TE mode at different regions along the LN coupler are shown in Fig. \ref{fig1}(b-d). As expected, the mode size becomes larger and more symmetric as light is transferred to the tapered slab layer such that nearly all optical power remains in the target eigenmode. Our optimal coupler design features a fiber-to-rib coupling efficiency as high as $88\%$.

It should be noted that the MFD we obtained here is not the largest possible but is chosen to match the mode size of the lensed fiber used. It is possible to interface optical fibers with larger optical modes by tapering the waveguides to a even narrower width. We calculated the MFD of different tip widths $w_2$ changing from 50 nm to 1 $\mu$m at the telecom wavelength as shown in Fig. \ref{fig2} (blue curve). The mode size at the chip facet decreases greatly as the taper becomes wider, due to the stronger light confinement (higher effective index). When the taper is wider than 0.8 $\mu$m, the mode diameter grows gradually and finally reaches a plateau. By changing the widths of tip, we can achieve 1.4-4.5 $\mu$m MFD at the edge of the chip. Furthermore, the implementation of our bilayer inverse taper successfully brings the effective refractive index of the rib waveguide mode from 1.9 down to a range between 1.44-1.62 (red curve in Fig. \ref{fig2}). Therefore, the index mismatch between lithium niobate photonic integrated devices and lensed fiber is well compensated.

\begin{figure}
	\centering
	\includegraphics[angle=0,width=0.5\textwidth]{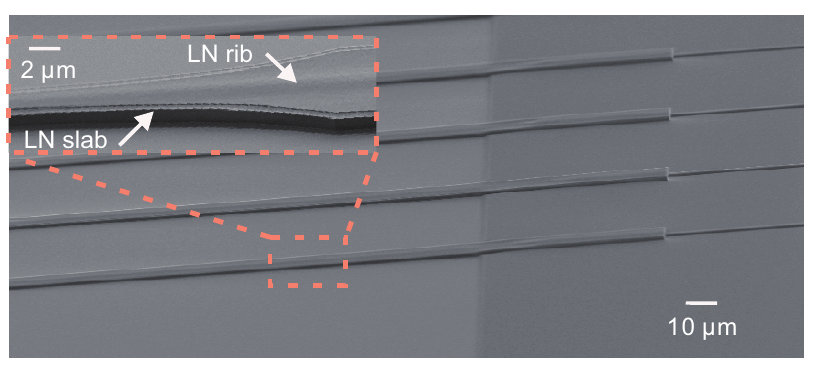}
	
	\caption{\label{fig3}\textbf{Scanning electron microscope (SEM) images of the fabricated bilayer mode size converter.} Inset: a close-up view of the bottom-layer taper. The top-layer waveguide is covered by e-beam resist after patterning the slab, so the contour is blurred. }
\end{figure}

We fabricated the converter using a commercially available x-cut LNOI substrate (NANOLN), where a LN thin film (600 nm thick) was bonded on top of silica (2 $\mu$m thick) on a silicon substrate handle (0.5 mm thick). The top-down fabrication process involves a two-step lithography and etching process of LNOI wafer. In the first step, we defined the patterns of the intended LN photonic integrated circuits and the top taper layer using electron beam lithography (EBL) and transferred the patterns 350 nm deep into LN thin film using reactive ion etching (RIE) with argon \cite{RN66}. The waveguides were then cladded by depositing 1 $\mu$m of SiO$_2$ using plasma-enhanced chemical vapor deposition (PECVD). Aligned photolithography and hydrofluoric acid (HF) wet etching were then employed to remove the cladding layer at the corresponding tapering regions, while the rest of the photonic integrated chip is protected from subsequent etching. After a second layer of aligned EBL and RIE, the second taper layer was defined and re-cladded with PECVD SiO$_2$. The facets of the final devices were etched to reduce roughness and to ensure good coupling from and to the optical fibers. Fig. \ref{fig3} displays a scanning electron microscope (SEM) image of the mode size converters we fabricated.

\begin{figure}
	\centering
	\includegraphics[angle=0,width=0.5\textwidth]{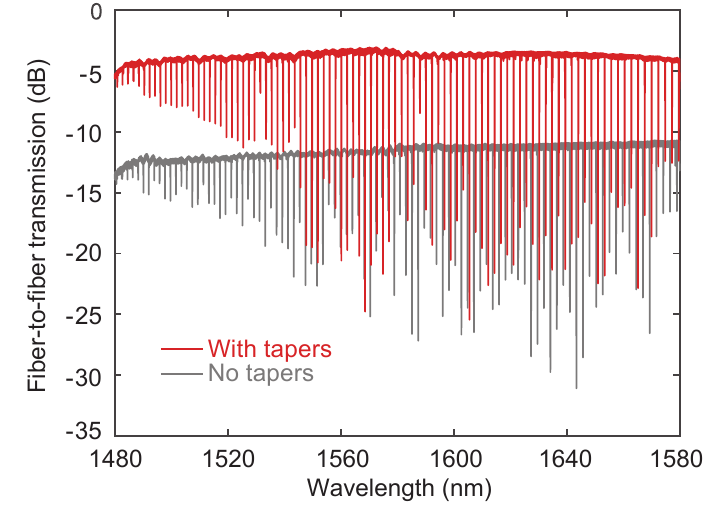}
	\caption{\label{fig4}\textbf{Calibrated transmission spectra.} The red and grey curves correspond to the spectra for LN rib waveguides coupled to ring resonators with and without couplers, showing fiber-to-fiber insertion losses of 14 dB and 3.4 dB, respectively. Dips correspond to microring cavity resonances, featuring $Q_{loaded} = 1.5 \times 10^6$.}
\end{figure}

We use a transmission setup to measure the fiber-to-fiber insertion losses of LN chips with and without mode size converters. A tunable telecom external cavity diode laser  (Santec-TSL 510) is used to excite the devices, and the transmitted light is collected and detected using an InGaAs photodetector  (Newport 1811). A piezo-controlled micropositioner is employed to precisely control the motion of the tapered-lensed fiber to achieve optimal coupling with the chip. To ensure TE input polarization and maximizes the transmitted laser light, a fiber polarization controller was placed after the laser. We also fabricated microring resonators on the chip to confirm that the transmission we observe are not from slab modes. Fig. \ref{fig4} displays the broadband transmission spectra of the on-chip LN mode size converter coupled to the tapered-lensed fiber, which exhibits a series of dips attributed to localized TE-polarized microring resonances (red curve). The coupling loss decreases from 11-14 dB (grey curve) to $\sim$ 3-5 dB (red curve) from 1480 nm to 1680 nm after introducing the mode size converter. Note that both waveguides, with and without mode size converter, are 4 mm long and are fabricated on the same chip to avoid possible variations between different fabrication runs. We experimentally confirm the low propagation loss of our waveguides \cite{RN66} using the measured quality factor ($Q$) of a nearly critically coupled microring resonator on the same chip as a reference ($Q_\textrm{loaded}\sim 1.5 \times 10^6$). The high quality factor indicates that the 4 mm waveguides have a propagation loss of $<$ 0.05 dB. Therefore, we conclude that the overall insertion loss that we measure is dominantly due to fiber-to-chip coupling. We carried out a series of measurements for devices with different slab tip widths, and found a minimum fiber-to-fiber insertion loss of 3.4 dB (\ref{fig4} red curve) for a tip width of 340 nm. In order to evaluate the robustness and compatibility of our devices for high power applications we tested our devices under high optical powers up to 1 W for continuous operation of more than 1 hour, and no damage to the facet and coupling region was observed.

\begin{figure*}
	\centering
	\includegraphics[angle=0,width=0.9\textwidth]{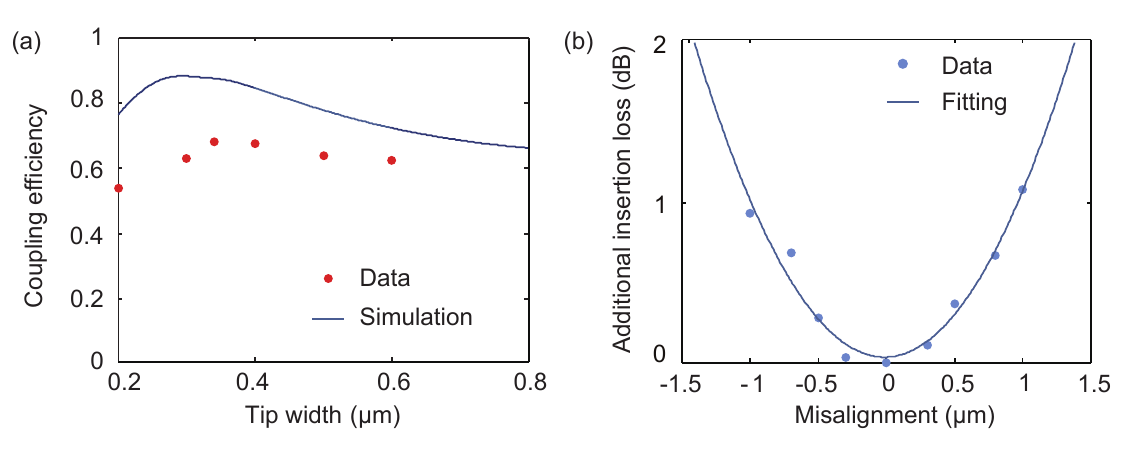}
	\caption{\label{fig5}\textbf{Measurement results.} \textbf{(a)} Simulated (blue curve) and measured (red dots) coupling efficiencies versus different tip widths of the tapered slab region, respectively. \textbf{(b)} Additional insertion loss of the bilayer nanotaper as a function of fiber-chip misalignment (TE mode).}
\end{figure*}

We compare the experimentally measured coupling efficiencies for different taper widths, and find them to be in good agreement with theoretical predictions (Fig.\ref{fig5}(a)). The difference in the experimental data and simulation may be attributed to misalignment between the two-tapers and/or non-ideal etch depths of rib waveguides. We also measured the tolerance of the optical fiber position for TE polarization. The input fiber was mounted on the motorized stage and scanned over 4 $\mu$m in the horizontal plane while the distance between the input fiber and chip facet is fixed at the confocal point of lensed fiber. The measured alignment tolerance for 1 dB excess loss is $\pm$ 1 $\mu$m in the transverse directions, as shown in Fig. \ref{fig5}(b). The alignment tolerance agrees with our estimate of the mode size.

In conclusion, we have experimentally demonstrated a monolithic bilayer mode size converter for efficient fiber coupling to LN nanophotonic waveguides. The mode size converter consists of two-layer taper, implemented using a two-step dry etching process, that gradually decreases the width of both rib and slab portions of LN waveguide. Using this approach, the total insertion loss could be reduced from 14 dB to 3.4 dB. The high coupling efficiency will enable optical packaging of LNOI devices, and opens the door for practical applications in LN integrated photonics. We note that our approach can be extended to coupling to standard, cleaved, optical fibers  (e.g. Corning SMF28, with the MFD of 10.4 $\pm$ 0.5 $\mu$m at 1550 nm) using high numerical aperture fiber (MFD$\sim$3.2 $\mu$m) as an intermediary. This method has been successfully employed to couple light efficiently from SMF28 to silicon waveguide with an overall coupling loss of less than 1.5 dB \cite{preble2015chip}.

This work is supported by: NSF Partnership for Innovation - Technology Translation (PFI-TT) program (Award IIP-1827720), China Scholarship Council (201706470030), City University of Hong Kong (Start-up Funds). Device fabrication is performed at the Harvard University Center for Nanoscale Systems, a member of the National Nanotechnology Coordinated Infrastructure Network, which is supported by the National Science Foundation under award number ECCS-1541959. 

We also thank Mengjie Yu for assistance in high optical power measurement, Linbo Shao, Christian Reimer, and Boris Desiatov for fruitful discussions and.

\bibliography{reference}

\end{document}